\documentclass[aps,pre,twocolumn,nofootinbib,floatfix]{revtex4}
\usepackage{amsmath}
\usepackage{epsfig}

\input{tcilatex}

\begin{document}

\title{Alignment and orientation of an absorbed dipole molecule}
\author{Y. Y. Liao, Y. N. Chen, and D. S. Chuu}
\affiliation{Department of Electrophysics, National Chiao-Tung University, Hsinchu 300,
Taiwan}
\date{\today }

\begin{abstract}
Half-cycle laser pulse is applied on an absorbed molecule to investigate its
alignment and orientation behavior. Crossover from field-free to hindered
rotation motion is observed by varying the angel of hindrance of potential
well. At small hindered angle, both alignment and orientation show
sinusoidal-like behavior because of the suppression of higher excited
states. However, mean orientation decreases monotonically as the hindered
angle is increased, while mean alignment displays a minimum point at certain
hindered angle. The reason is attributed to the symmetry of wavefunction and
can be explained well by analyzing the coefficients of eigenstates.

PACS: 73.20.Hb, 33.55.Be, 33.20.Sn
\end{abstract}

\maketitle

\address{Department of Electrophysics, National Chiao Tung University,
Hsinchu 300, Taiwan}





Alignment and orientation of molecules are important in the investigations
of stereodynamics \cite{1}, surface catalysis \cite{2}, molecular focusing %
\cite{3}, and nanoscale design \cite{4}. The alignment scheme has been
demonstrated both in adiabatic and nonadiabatic regimes. A strong laser
pulse can adiabatically create pendular states, and the molecular axis is
aligned in parallel to the direction of field polarization. The molecule
goes back to its initial condition after the laser pulse is switched of, and
the alignment can no longer be observed again\cite{5}. To achieve adiabatic
alignment, the duration of laser pulse must be longer than the rotational
period. However, an ultrashort laser pulse with several cycles is also
observed to induce a field-free alignment providing the duration of laser
pulse is smaller than the rotational period. In this limit, the alignment
occurs periodically in time as long as the coherence of the process is
preserved \cite{6}. On the other hand, a femtosecond laser pulse is found to
be able to generate field-free orientations \cite{7}. The dipole molecule,
kicked by an impulsive pulse, will tend to orient in the direction of laser
polarization. Laser-indued molecular orientations have been demonstrated in
several experiments \cite{8,9,10}.

Recently, the rotational motion of a molecule interacting with a solid
surface has attracted increasing interest. It is known that molecules can be
desorbed by applying UV laser beam along the surface direction, and the
quadrupole is a measure of the rotational alignment \cite{11}. To understand
molecular-surface interaction, Gadzuk and his co-workers \cite{12} proposed
an infinite-conical-well model, in which the adsorbed molecule is only
allowed to rotate within the well region. Shih \textit{et al}. further
proposed a finite-conical-well model to generalize the study of a finite
hindrance \cite{13}. Their results showed that the rotational states of an
adsorbed dipole molecule in an external electric field exhibit interesting
behaviors, and theoretical calculation of the quadrupole moment based on
finite-conical-well model is in agreement with the experimental data \cite%
{14}. These findings may be great useful for understanding the surface
reaction.

In the present work we present a detailed investigations on the rotational
motion of a polar diatomic molecule, which is confined by a hindering
conical-well. Different well-dependent signatures between the alignment and
orientation of the hindered molecule under an ultrashort laser pulse are
pointed out for the first time. Crossover from field-free to hindered
rotation is also observed by varying the hindered angle of potential well.
These make our results promising and may be useful in understanding the
molecule-surface interactions. 
\begin{figure}[th]
\includegraphics[width=7.5cm]{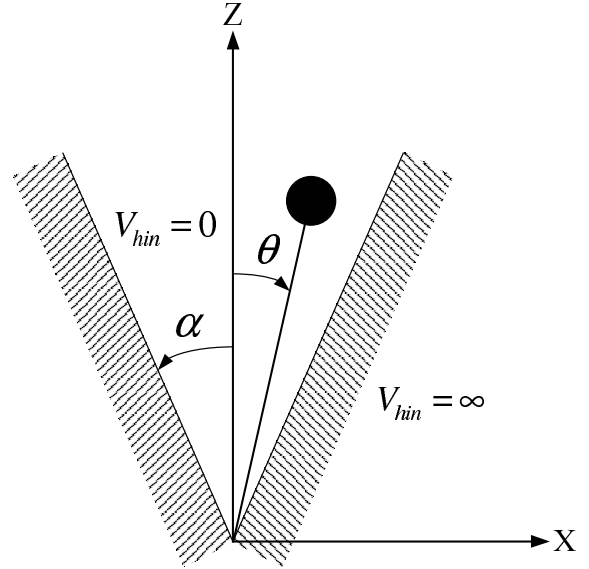}
\caption{Schematic view of the hindered rotor.}
\end{figure}

Consider now a laser pulse polarizing in z-direction interacts with the
hindered molecule as shown in Fig.1. The model Hamiltonian can be written as%
\begin{equation}
H=-\frac{\hbar ^{2}}{2I}J^{2}+V_{hin}(\theta ,\phi )+H_{I},
\end{equation}%
where $J^{2}$\ and $I$ \ are angular momentum and momentum of inertia of the
molecule. The rotational constant $B$ is set equal to $\frac{\hbar ^{2}}{2I}$%
. It is reasonable to assume that the surface potential $V_{hin}(\theta
,\phi )$ is independent of $\phi $ since previous calculations have shown
that its dependence on $\phi $ is weaker than that on $\theta $ \cite%
{15,16,17}. Therefore, in the vertical absorbed configuration, the surface
potential can be written as \cite{12}%
\begin{equation}
V_{hin}(\theta )=\left\{ \QATOP{0,\text{ }0\leq \theta \leq \alpha }{\infty ,%
\text{ }\alpha <\theta \leq \pi }\right. ,
\end{equation}%
where $\alpha $ is the hindered angle of the conical well. In Eq. (1), $%
H_{I} $ describes the interaction between the dipole moments (permanent and
induced) and laser field:

\begin{eqnarray}
H_{I} &=&-\mu E\left( t\right) \cos \theta  \notag \\
&&-\frac{1}{2}E^{2}\left( t\right) ((\alpha _{\parallel }-\alpha _{\perp
})\cos ^{2}\theta +\alpha _{\perp }),
\end{eqnarray}%
where $\mu $ is the dipole moment. The components of the polarizability $%
\alpha _{\parallel }$ and $\alpha _{\perp }$ are parallel and perpendicular
to the molecular axis, respectively The laser field in our consideration is
a Gaussian shape: $E\left( t\right) =Ee^{-\frac{(t-to)^{2}}{\sigma ^{2}}%
}\cos \left( \omega t\right) ,$ where $E$ is the field strength and $\omega $
is the laser frequency. To solve time-dependent Schr\"{o}dinger equation,
the wavefunction is expanded in terms of a series of eignfunctions as:%
\begin{equation}
\Psi _{l,m}=\sum c_{l,m}\left( t\right) \psi _{l,m}\left( \theta ,\phi
\right) ,
\end{equation}%
where $c_{l,m}\left( t\right) $ is time-dependent coefficients corresponding
to the quantum numbers $\left( l,m\right) $. For the vertical adsorbed
configuration, the wavefunction can be written as 
\begin{equation}
\psi _{l,m}\left( \theta ,\phi \right) =\left\{ \QATOP{A_{l,m}P_{\nu
_{l,m}}^{\left| m\right| }(\cos \theta )\frac{\exp (im\phi )}{\sqrt{2\pi }},%
\text{ }0\leq \theta \leq \alpha }{0\text{ \ \ \ \ \ \ \ \ \ \ \ \ \ \ \ \ \
\ \ \ \ \ \ \ \ \ \ \ \ },\alpha <\theta \leq \pi }\right. ,
\end{equation}%
where $A_{l,m}$ is the normalization constant and $P_{\upsilon
_{l,m}}^{\left| m\right| }$ is associated Legendre function of arbitrary
order. In above equations, the molecular rotational energy can be expressed
as 
\begin{equation}
\epsilon _{l,m}=\nu _{l,m}(\nu _{l,m}+1)B.
\end{equation}%
In order to determine $\nu _{l,m}$, one has to match the boundary condition%
\begin{equation}
P_{\nu _{l,m}}^{\left| m\right| }\left( \cos \alpha \right) =0.
\end{equation}%
After determining the coefficients $c_{l,m}\left( t\right) $, the
orientation $\left\langle \cos \theta \right\rangle $ and alignment $%
\left\langle \cos ^{2}\theta \right\rangle $ can be carried out immediately.

We choose ICl as our model molecule, whose dipole moment $\mu $ = 1.24
Debye, rotational constant $B$\ =0.114 cm$^{-1}$, polarizability components $%
\alpha _{\parallel }\approx 18$ \AA $^{3}$ and $\alpha _{\perp }\approx 9$ 
\AA $^{3}$. The peak intensity and frequency of laser pulse is about $%
5\times 10^{11}$ W/cm$^{2}$ and $210$ cm$^{-1}$, respectively. For
simplicity (zero-temperature case), the rotor is assumed in ground state
initially, i.e. $c_{0,0}\left( t=0\right) =1$. Besides, in order to keep the
simulations promising, the highest quantum number for numerical calculations
is $l=15$, such that the results are convergent and the precision is to the
order of $10^{-7}$.

The solid lines in the insets of Fig. 2 show the dependence of the alignment
on hindered angle $\alpha .$ For $\alpha =60^{\circ }$, \ sinusoidal-like
behavior is presented, and the alignment ranges from 0.63 to 0.91. As the
hindered angle increases, the curves become more and more complicated and
gradually approach the free rotor limit as shown in the insets of Fig. 2 (b)
($\alpha =120^{\circ }$) and 2 (c) ($\alpha =180^{\circ }$). This can be
understood well by studying the populations $\left| c_{l,m}\right| ^{2}$ of
low-lying states. In the regime of small hindered angle, there is little
chance for electron to populate in higher excited states since the shrinking
of the conical-well angle causes the increasing of energy spacings. 
\begin{figure}[th]
\includegraphics[width=7.5cm]{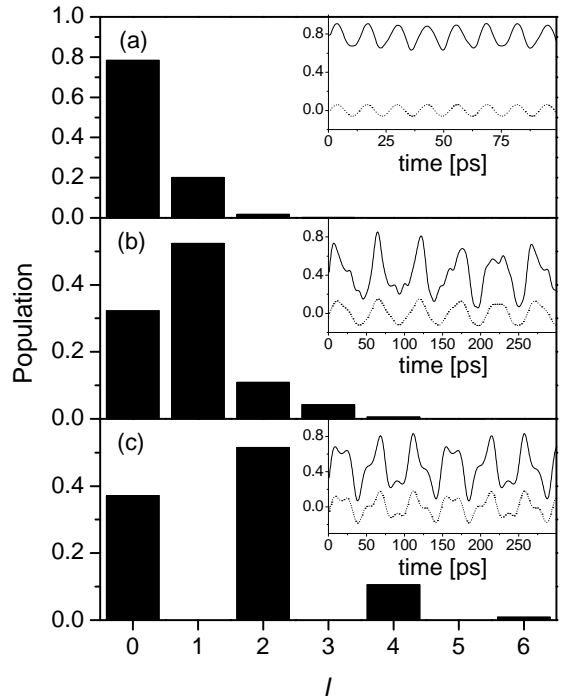}
\caption{The populations of the states $\left( l,m=0\right) $ for different
hindered angles: (a) $\protect\alpha =60^{0}$, (b) $\protect\alpha =120^{0}$%
, (c) $\protect\alpha =180^{0}$. The insets show the corresponding
alignments (solid lines) and the factors $\protect\underset{l\neq l^{\prime }%
}{\sum }\left\langle \protect\psi _{l^{\prime },m^{\prime }}\left| \cos ^{2}%
\protect\theta \right| \protect\psi _{l,m}\right\rangle $ (dotted lines).}
\end{figure}

One also notes that the populations of a hindered molecule for $\alpha
=60^{\circ }$ and $120^{\circ }$, shown in Fig. 2 (a) and (b), mainly
compose of $l=0$, $1$ and $2$ states, while the population of a free rotor
is composed of $l=0,$ $2,4$ states. The underlying physics comes from the
reason that $\left\langle \psi _{l^{\prime },m^{\prime }}\left| \cos
^{2}\theta \right| \psi _{l,m}\right\rangle $ is non-zero for all $l$ and $%
l^{\prime }$ values in the case of hindered rotation. But it is zero in free
rotor limit except for $l=l^{\prime }$ or $l=l^{\prime }\pm 2$. The dotted
lines in the insets represent the first two contributions of the factors $%
\underset{l\neq l^{\prime }}{\sum }\left\langle \psi _{l^{\prime },m^{\prime
}}\left| \cos ^{2}\theta \right| \psi _{l,m}\right\rangle $ from low-lying
states, i.e. summing the largest value of the off-diagonal term $%
\left\langle \psi _{l^{\prime },m^{\prime }}\left| \cos ^{2}\theta \right|
\psi _{l,m}\right\rangle $. As can be seen, the populations for small
hindered angle are mainly distributed on lower states since the main
oscillation frequency of the curve (dotted lines) is roughly equal to that
from whole contributions (solid lines). 
\begin{figure}[th]
\includegraphics[width=7.5cm]{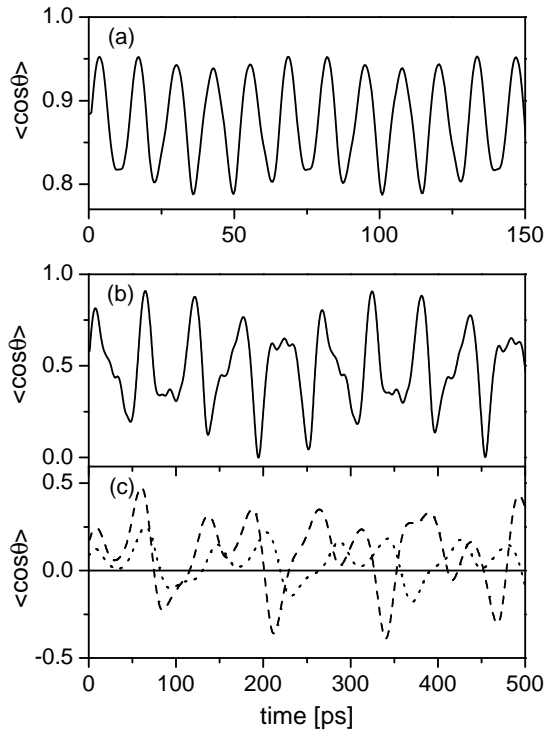}
\caption{The orientations $\left\langle \cos \protect\theta \right\rangle $
(solid lines) of a hindered molecule confined by infinite conical-well for
different hindered angles: (a) $\protect\alpha =60^{0}$, (b) $\protect\alpha %
=120^{0}$, (c) $\protect\alpha =175^{0}$. The dashed and dotted lines in (c)
correspond to different potential barrier height, i.e. $V_{0}=\infty $ and
100, respectively. }
\end{figure}

Let us now turn our attention to the case of orientation. After applying a
short pulse laser, the orientation $\left\langle \cos \theta \right\rangle $
of a hindered molecule ($\alpha =60^{\circ }$) oscillates sinusoidally with
time as shown in Fig. 3 (a). The value of $\left\langle \cos \theta
\right\rangle $ is always positive because the rotational wavefunction is
compressed heavily. As the hindered angle $\alpha $ becomes larger, the
oscillation frequency also decreases as shown in Fig. 3 (b). These
signatures are quite close to that of the alignment. We then conclude that
even at larger hindered angle ($\alpha =120^{0}$) the role of hindered
potential still overwhelms the laser pulse, or the value of $\left\langle
\cos \theta \right\rangle $ should not always be positive.

Fig. 3 (c) represents results of orientations in infinite ($V_{0}=\infty $)
or finite ($V_{0}=100$) conical-well potential for $\alpha =175^{0}$. Dashed
and dotted lines correspond to $V_{0}=\infty $ and $100$, respectively. As
can be seen, the effect of laser pulse is obvious because negative value
appears. Comparing the results with the free orientation \cite{7}, the
angular distributions for finite well are more isotropic since the wave
functions can penetrate into the conical-barrier. 
\begin{figure}[h]
\includegraphics[width=8cm]{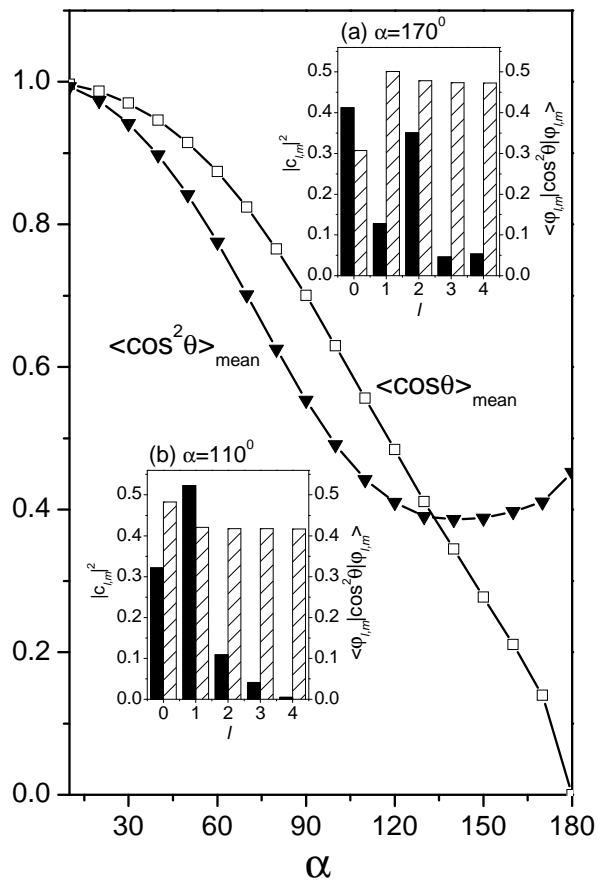}
\caption{The mean orientation $\left\langle \cos \protect\theta %
\right\rangle _{\text{mean}}$ and alignment $\left\langle \cos ^{2}\protect%
\theta \right\rangle _{\text{mean}}$ in infinite conical-well. The insets
show the populations $\left| c_{l,m}\right| ^{2}$ (fulled bar) and factors $%
\left\langle \protect\psi _{l,m}\left| \cos ^{2}\protect\theta \right| 
\protect\psi _{l,m}\right\rangle $ (sparse bar). Insets (a) and (b)
correspond to $\protect\alpha =110^{0}$ and $\protect\alpha =170^{0},$
respectively.}
\end{figure}

Further analysis shows that components of orientation $\left\langle \cos
\theta \right\rangle $ or alignment $\left\langle \cos ^{2}\theta
\right\rangle $ can be divided into two parts: diagonal and nondiagonal
terms. The nondiagonal term represents the variations of these curves such
as those in the insets of Fig. 2. These variations with time are determined
by the phase difference coming from various energy levels. To see the
contributions from diagonal terms, we evaluate the time-averaged orientation
and alignment. In this case, the nondiagonal values will be averaged out,
and only contributions from diagonal terms exit. Fig. 4 shows the mean
orientation and alignment as a function of hindered angle. As $\alpha $
increases, the mean orientation decreases monotonically from 1 to 0. This is
because the mean orientation is determined by $\left| c_{l,m}\right| ^{2}$
and $\left\langle \psi _{l,m}\left| \cos \theta \right| \psi
_{l,m}\right\rangle $. For a larger angle $\alpha ,$ the populations $\left|
c_{l,m}\right| ^{2}$ mainly compose of $l=0,2$, $4$ states. But the value $%
\left\langle \psi _{l,m}\left| \cos \theta \right| \psi _{l,m}\right\rangle $
is governed by the selection rule: $l=l^{\prime }+1$. Thus the net effect is
the shrinking of the mean orientation in large angle limit.

Contrary to orientation, the mean alignment shows a quite different feature.
The value of $\left\langle \cos ^{2}\theta \right\rangle $ first decreases
as $\alpha $ increases. However, it reaches a minimum point about for $%
\alpha =140^{0}$. From the insets of Fig. 4, we know that the values of $%
\left( \left\langle \psi _{l,m}\left| \cos ^{2}\theta \right| \psi
_{l,m}\right\rangle \right) $ do not depend significantly on $\alpha $.
Therefore, the decrease of $\left\langle \cos ^{2}\theta \right\rangle $
comes from the decreasing tendency of the population $\left|
c_{l=1,m}\right| ^{2}$, while its increasing behavior is caused by other two
populations $\left| c_{l=0,m}\right| ^{2}$ and $\left| c_{l=2,m}\right| ^{2}$%
. Competition between these two effects results in a minimum point.

A few remarks about experimental verifications should be mentioned here. The
degree of alignment can be measured with the techniques of the femtosecond
photodissociation spectroscopy and the ion imaging \cite{8}. The alignment
is probed by breaking the molecular bond and subsequently measuring the
direction of the photo-fragments by a mass selective position sensitive ion
detector. In contrast to alignment, the orientation is probed by Coulomb
exploding the molecules with a femtosecond laser pulse \cite{9}. By
detecting the fragment ions with the time-of-flight mass spectrometer, a
significant asymmetry should be observed in the signal magnitudes of the
forward and the backward fragments. Under proper arrangements, orientation
and alignment of an absorbed molecule may be examined by these spectroscopic
technologies.

In conclusion, we have shown that a short laser pulse can induce alignment
and orientation of a hindered molecule. The hindered angle of hindered
potential well plays a key role on the molecular alignment and orientation.
Crossover from field-free rotation to hindered one can be observed by
varying the hindered angle of the potential well. Time averaged alignment
and orientation are investigated thoroughly to understand the difference
between these two quantities.

This work is supported partially by the National Science Council, Taiwan
under the grant number NSC 92-2120-M-009-010.

\bigskip

\end{document}